\documentclass[twocolumn]{revtex4}
\usepackage{times}
\usepackage{color}
\usepackage{amsmath}
\usepackage{amssymb}
\usepackage{graphicx}% Include figure files
\def\lesssim{\mathrel{\mathpalette\vereq<}}

\usepackage{color}

\newcommand{\be}{\begin{equation}}
\newcommand{\ee}{\end{equation}}
\newcommand{\bea}{\begin{eqnarray}}
\newcommand{\eea}{\end{eqnarray}}
\newcommand{\bc}{\begin{center}}
\newcommand{\ec}{\end{center}}

\begin{document}
\title{Peccei-Quinn axions from frequency dependence radiation dimming}
\author{Raul Jimenez$^{1}$, Carlos Pe\~na-Garay$^{2}$, Licia Verde$^{1}$ }
\affiliation{
$^{1}$ ICREA \& ICC, University of Barcelona (IEEC-UB), Spain.\\
$^{2}$ Instituto de F\'isica Corpuscular (CSIC-UVEG), Val\`encia, Spain.\\
}
\date{\today}
\begin{abstract}
 We explore how the Peccei-Quinn (PQ) axion parameter space can be constrained by the frequency-dependence dimming of  radiation  from astrophysical objects. To do so we perform accurate calculations of photon-axion conversion in the presence of a variable magnetic field. We  propose several tests where the PQ axion parameter space can be explored with current and future astronomical surveys: the observed spectra of isolated neutron stars, occultations of background objects by  white dwarfs and neutron stars, the light curves of eclipsing binaries containing a white dwarf. We find that 
 the lack of dimming of the light curve of a detached eclipsing white dwarf binary recently observed, leads to relevant constraints on the photon-axion
 conversion. Current surveys designed for Earth-like planet searches are well matched to strengthen  and improve the constraints on the PQ axion using astrophysical objects radiation dimming.
\end{abstract}
\pacs{?}
\maketitle

%%%%%%%%%%%%%%%%%%%%%%%%%%%%%%%%%%%%%%%%%%%%%%%%%%%%%%%%%%%%%%%%%%%%%%%

%\section{Introduction}

One of the most attractive solutions to the strong CP problem is to introduce the Peccei-Quinn (PQ) symmetry\cite{PQ}. This anomalous global symmetry, leads to a pseudo-Nambu-Goldstone boson, the axion, whose vacuum expectation value permits to solve the problem. Most searches for this pseudoscalar particle are currently being performed in laboratory experiments and in astrophysical objects by their generic (though model 
dependent) coupling to two photons \cite{raffelt}. 

No PQ axion has been observed yet, which narrows the allowed region in parameter space defined by the mass of the axion versus its coupling to photons. In the presence of a magnetic field, a photon with polarization contained on the plane defined by the external magnetic field and the photon propagation direction, can be converted into an axion and viceversa. This leads, among others, to possible phenomena like extra release of energy in stars or  appearance of x-rays in strong magnetic fields pointing to the Sun.

 It has long been recognized that neutron stars (NS) are excellent laboratories to detect axions via the conversion effect, because the sensitivity 
 to the axion-photon coupling grows with the magnetic field strength \cite{Chelouche:2008ta}. 
 Big efforts have been focused on axions as dark matter candidates \cite{reviewaxionsdm}, but axions may exist and not be the dominant component of the dark matter. On the other hand, independently of the fact that axion are the dark matter, the PQ axions could leave their signature
   in the photon emission of astrophysical objects such as NS and white dwarfs (WD).
 
Here, we explore the possibility of detecting PQ axions by the frequency-dependent dimming of light of astrophysical objects 
 crossing strong magnetic fields.  In particular, we consider three configurations: a)isolated NS, b) occultation of background objects by NS and WD, and c) ``eclipsing" binary systems containing a WD. We find the last case to be the more promising, and also least dependent on astrophysical assumptions. Currently on-going surveys driven by the search of substellar companions of WD, will provide a large sample suitable for this purpose  and the first WD-WD eclipsing binary has already been observed \cite{WD-WD}. 
We show that the data of  \cite{WD-WD}  already place significant bounds on the photon-axion conversion. These bounds will be further tested by the $\sim$ hundred systems expected to be found by Kepler\cite{Kepler}.
 
%\section{Photon Conversion into axions}
%
Photons with polarization in the plane formed by the magnetic field and their propagation directions, $\wp_{\parallel}$, 
can be converted into axions. The probability $P_{\gamma\rightarrow \phi}(L)$ of a photon converting 
into an axion after travelling a distance $L$ in a constant, coherent magnetic field $B$ 
is given by\cite{rs,donglai}:
\be\label{PofL}
P_{\gamma\rightarrow \phi}(L)=\sin^2(2\psi) \sin^2\left(\frac{\Delta k}{2}L\right)
\ee
where
\be%\label{tanofpsi}
\tan 2\psi=\frac{2\Delta_{\rm M}}{\Delta_{\rm a}-\Delta_{\rm \gamma}} \,,\,\,\,\,\Delta_{\rm M}=\frac{B}{2 M} \,,\,\,\,\,\Delta_{\rm a}=\frac{m_{\rm eff}^2}{2\omega}\,,
\ee
\be%\label{DeltaofL}
\Delta_{\gamma}=\frac{q(B) \omega}{2}\,,\,\,\,\, \Delta k=\sqrt{(\Delta_{\rm a}-\Delta_{\gamma})^2 + 4 \Delta_{\rm M}^2} \,.
%\ee
%with
%\be\label{Deltas}
%\Delta_{\rm a}=\frac{m_{\rm eff}^2}{2\omega} ,
%\Delta_{\gamma}=\frac{q(B) \omega}{2},
%\Delta_{\rm M}=\frac{B}{2 M} \,.
\ee
Here, $\omega$ is the photon frequency, $M\simeq 2\pi f_a \alpha^{-1}$ 
the coupling energy scale ($f_a$ being the axion decay constant), $\alpha$ the fine structure constant and 
$m_{\rm eff}$ the effective mass:
%\be\label{meff}
$m_{\rm eff}^2=|m_a^2-\omega_{\rm P}^2| \,, $
%\ee 
%where 
with $\omega_{\rm P}$ the plasma frequency of the medium (which can be neglected for our purposes) and $m_a$ the axion mass.   Note that $M$ is related to the axion-photon coupling $g_{a\gamma\gamma}$ by $M\simeq 1/g_{a\gamma\gamma}$. The connection between the PQ mass scale and the photon-axion coupling is model-dependent, and particle physics constructions \cite{models} lead to a  factor  of order unity relating the two  independent couplings.
The function $q(B)$ can be approximated (for $b$ $\ll$ 1 and $b \gg$ 1) by the fitting formula \cite{Potekhin:2005ve}
\be\label{qB}
q(B)=\frac{7 \alpha b^2}{45 \pi} \frac{1 + 1.2 b}{1+1.33 b + 0.56 b^2} 
\,,
\ee
where $b$ is the magnetic field normalized to the critical QED field strength 
of  $4.414 \times 10^{13}$ G.

However,  Eq. (\ref{PofL}) does not reproduce the correct physics in most astrophysical situations, where the magnetic field varies 
along the photon path. For the astrophysical objects of our interest, the magnetic field is approximately dipolar (the field decreases as 
distance to the third power) and a correct treatment of the inhomogeneous magnetic field is required. 

Assuming that the magnetic field  is constant over distances 
comparable to the photon or axion wavelenghts, the evolution of the photon and axion amplitudes is given by solving the 
Schrodinger-like equation with Hamiltonian given by 
\begin{equation}
H = \begin{pmatrix} \Delta_a & \Delta_M \\ \Delta_M & \Delta_\gamma \end{pmatrix} \,.
\end{equation}
We obtain the photon-axion conversion by computing numerically the 
total evolution operator as the product of evolution operators in thin slices (1km width) of constant magnetic field, 
\begin{equation}
{\cal U} = \exp(-iHt) = \prod_j \exp[-i H(t_j) \delta t_j] \,.
\end{equation}
The total time, $t$, from production to detection  is subdivided in intervals $\delta t_j$ corresponding to $c \delta t_j=1$ km.

%\section{Astrophysical sources considered}
%{\bf Astrophysical sources considered}\\
We consider three possibilities based on the requirements of celestial objects to have large magnetic fields and weak dependence on 
complex astrophysical details:

a) The peculiar features of the XDINS spectra. These isolated NS (e.g., RXJ0720.4-3125, RXJ1856+5-3754, 
and RXJ1605.3+3249 \cite{xdins}) have been observed in the eV range and the near keV region. A single blackbody spectrum fits 
well the UV region (with temperatures $kT$=$75, 58, 115$ eV respectively), but would yield a too high hard X-ray photon flux. With the chosen temperatures, dimming factors of roughly $4, 7$ and $25$ respectively, are needed to fit the high-frequency 
(``X-ray dimmed") data. Simulations of the NS magnetic structure predict the photon polarization as a function of the frequency, 
changing from fully perpendicular, $\wp_{\perp}$, below  few $100$ eV to fully $\wp_{\parallel}$ at  higher frequencies  (\cite{donglai} 
and refs. therein); the frequency range where this transition happens is  model-dependent. 

b) ``Occultation" of background sources by NS and WD. If the light of a background source (star or galaxy) passes through 
the magnetic field of a NS or WD, it can be dimmed by photon-axion conversion. The relevant radius for a significant effect is that of maximum conversion, which is $\lesssim 10^3$ ($10^4$) km for NS (WD). This is the relevant input  for the cross section calculation quantifying the expected frequency of events.% per year.

c) ``Eclipses" in binary systems. This idea was originally explored for the double pulsar system detected in 2003 \cite{doublepulsar}. Unfortunately 
the number of binaries with NS is too low to be a good candidate to test the conversion into axions. We concentrate here in more conspicuous eclipsing binary systems where one of the companions is a WD and the other is a WD or a main sequence (MS) star. The Kepler telescope\cite{Kepler}, aiming at exoplanet transits, should be observing about a thousand of eclipsing WD-MS \cite{farmer}, where about 10\% of the WD have magnetic fields larger than 10$^6$ G.

\begin{figure}[t]
  \begin{center}
    \includegraphics[width=.75\columnwidth]{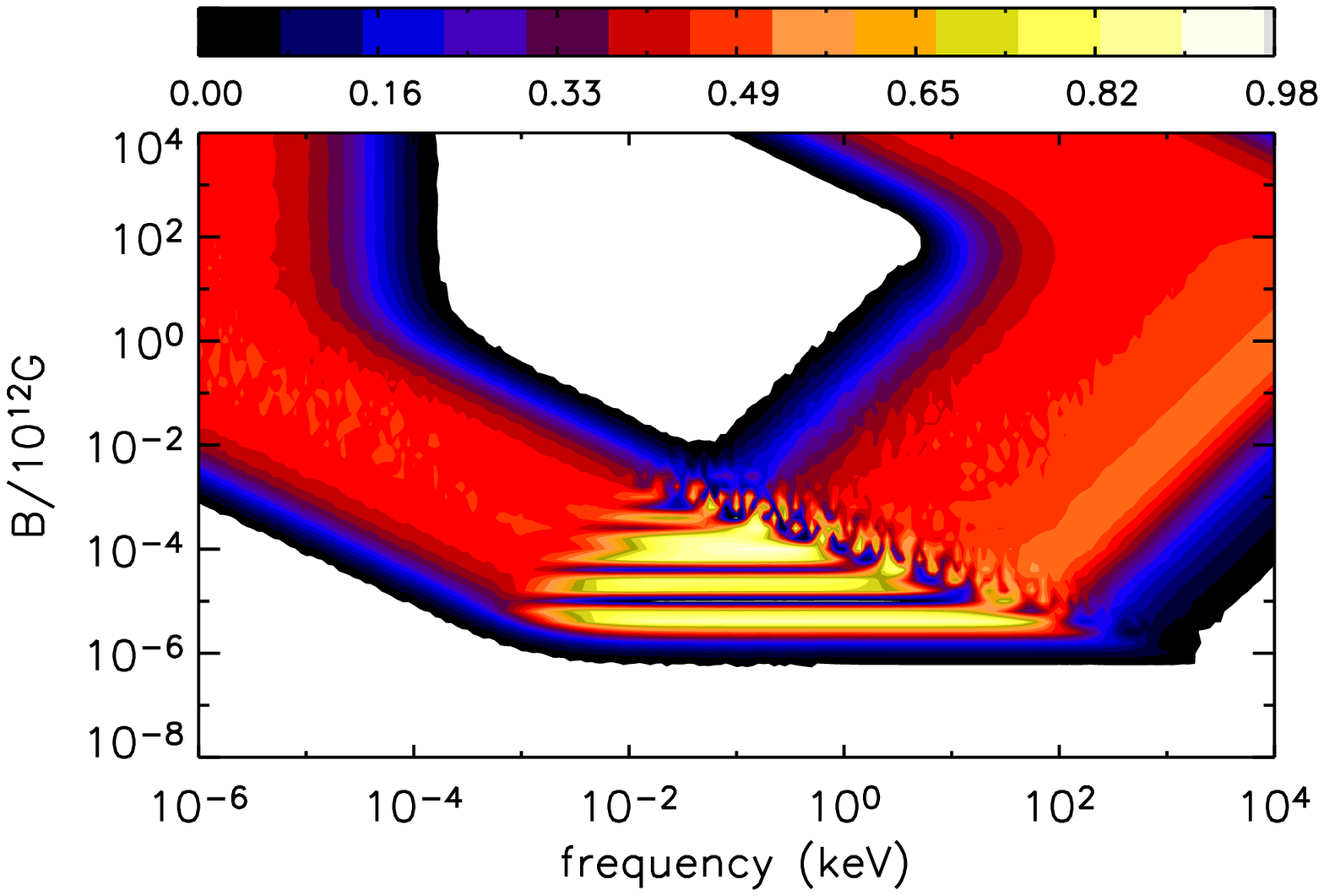}
    \includegraphics[width=.75\columnwidth]{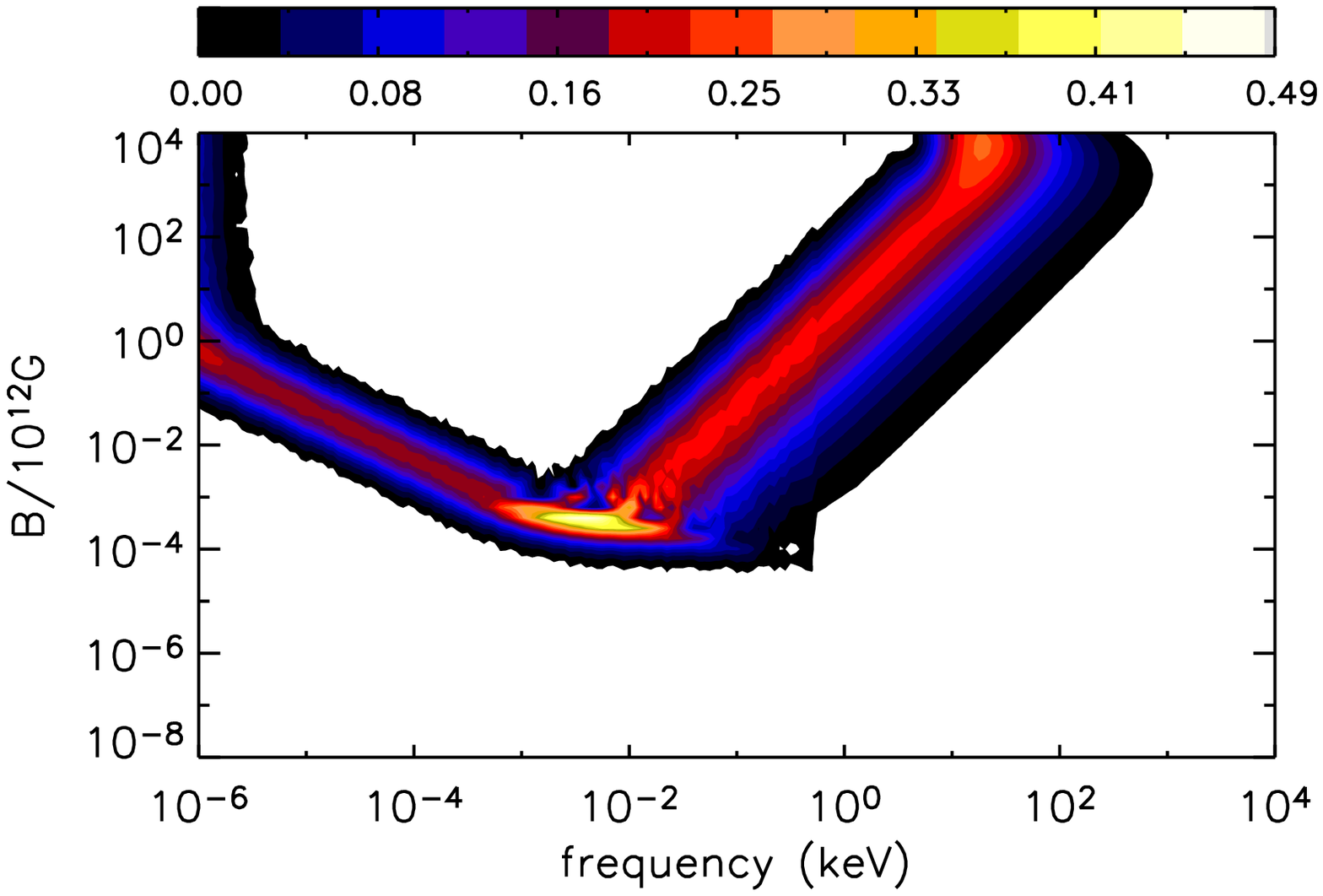}
    \caption{\label{fig:NSemission}  Probability conversion of $\wp_{\parallel}$ photons into axions for photons  exiting a NS dipolar magnetic field as a function of the star's magnetic field strength and photon energy.  The $y$-axis  denotes the NS surface magnetic field and we adopt a typical NS radius of $10$km.  We consider $m_a$=10$^{ -5}$ eV   and $M=10^7$($10^9$) GeV in the top (bottom) figure.}
    \label{fig:prob1}
  \end{center}
\end{figure}    

Let us begin by discussing  the probability of conversion of photons into axions produced on the surface of a source. We assume a pure beam
of $\wp_{\parallel}$ photons produced in the presence of a large magnetic field; the photon-axion system evolves outward towards weaker 
magnetic filed, where the photons are observed.  We show in Fig. \ref{fig:prob1} the dependence of the probability conversion 
into axions for a NS photons and for two illustrative $M$ values as a function of surface magnetic field and photon frequency. We consider a typical NS radius of $10$ km;  recall that typical surface magnetic fields of NS range from $10^6$ G  to $10^{13}$G.
In this calculation we account for the fact that the photon path is a geodesic (due to the high curvature of space time around the NS most light paths are not purely radial), but this has negligible effect on the final results. In fact most conversion happens at some distance for the star's surface (where the ratio of   crossing terms to the diagonal terms  in the Hamiltonian is maximal). Note that there are regions  with large  (90\% or higher) suppression which are wider for lower coupling energy scales.

Note the frequency-dependence in Fig.~\ref{fig:prob1}, which generally leads 
to a larger dimming effect at high frequencies. White regions of the plot correspond to zero conversion efficiency. Beside the frequency-dependent conversion, a novel remarkable feature is found in our analysis: the fact the maximum of the conversion appears at weaker  surface magnetic fields than the equipartition value. This implies that astrophysical systems with weaker magnetic fields than a NS can be suitable to explore photon-axion conversion. One excellent candidate for harboring such magnetic fields is a WD, about 10\% of them have magnetic field strengths of about or above $10^6 G$. The photon-axion adiabatic conversion is characterized by the mixing angle $\psi$ in the region of photon production, which is small for the weaker magnetic fields. However, for the objects we consider, it is  the non-adiabaticity of the photon-axion conversion  that plays a key role. The physical explanation of the net-effect is the non-adiabaticity: 
the polarization term is much smaller than the axion term,  the mixing angle $\psi$  changes  rapidly (as r$^{-3}$) along the photon path and the conversion happens in the region of maximum 
non-adiabaticity.  

\begin{figure}
  \begin{center}
    \includegraphics[width=.75\columnwidth]{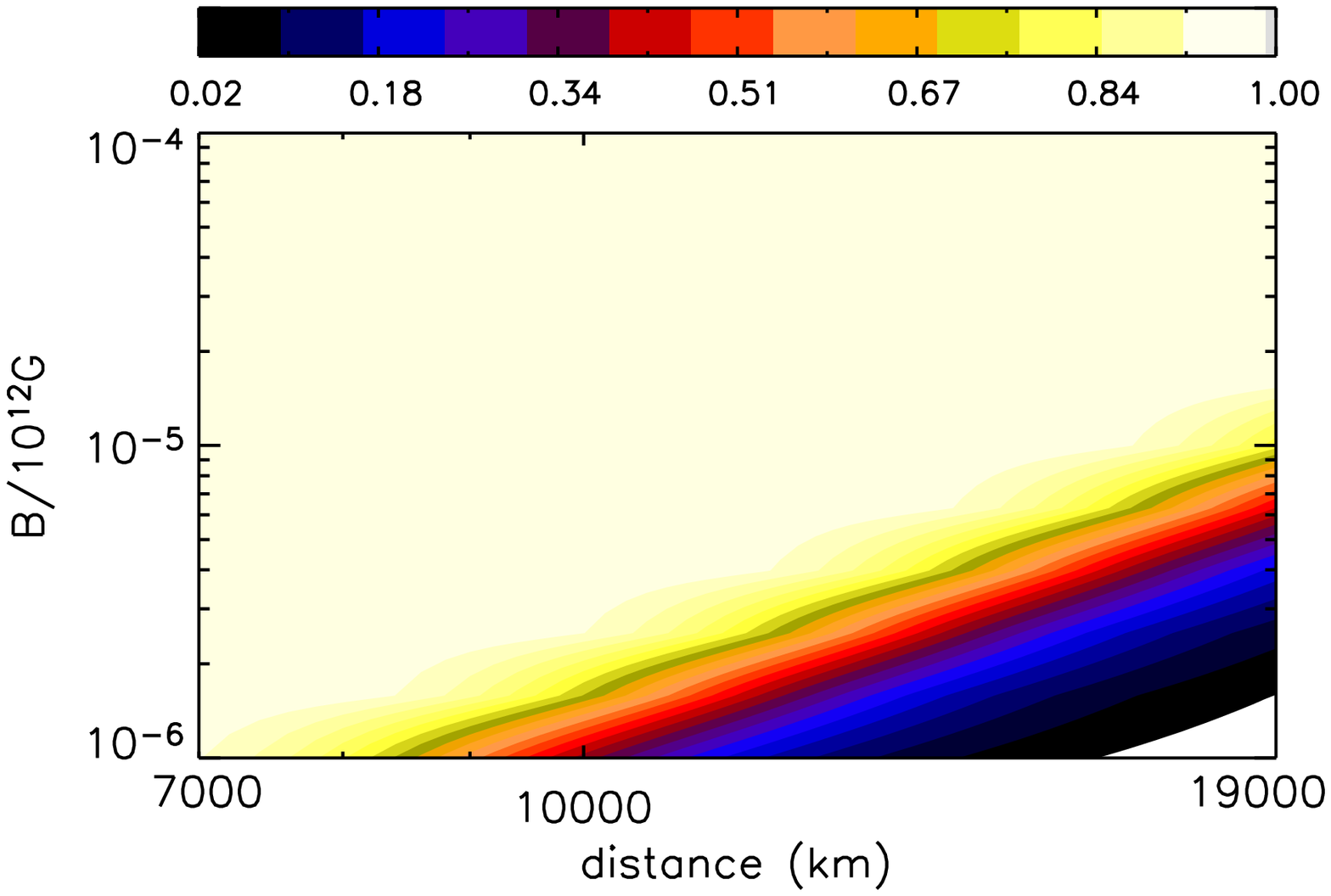}
    \includegraphics[width=.75\columnwidth]{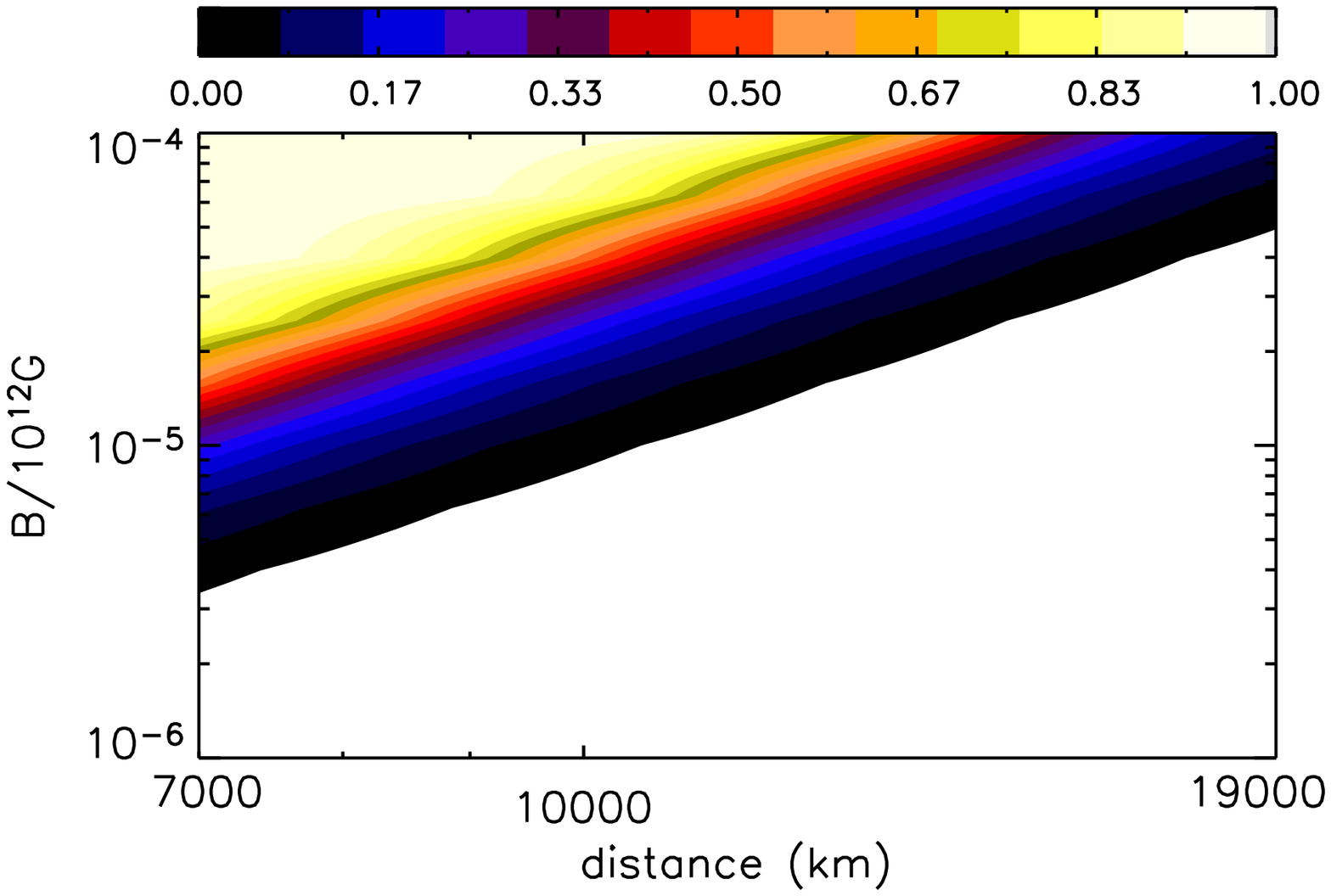}
    \caption{\label{fig:WDcrossing} Probability conversion of $\wp_{\parallel}$ photons into axion by crossing a star's dipolar magnetic field as a function of surface magnetic field strength and  distance to the center of the star.  We adopt a typical WD radius of $7000$km. The probability regions are  weakly dependent on the photon energy in the IR-VIS-UV region.  We consider $0.3$ eV photon energy, $m_a=10^{ -5}$ eV   and $M=10^7$(5$\times$10$^8$) GeV in the top (bottom) figure.}
    \label{fig:prob2}
  \end{center}
\end{figure}

One important implication of this finding is that, while external photons crossing adiabatically a magnetic field do not lead to a net conversion into axions, the non-adiabatic conversion opens new avenues for the exploration of photon dimming. Therefore, we explore the probability of conversion of photons into axions crossing the magnetic field of a foreground star.  In this case photons travel through an increasing magnetic field at first and later  decreasing. The geometry of the magnetic field  crossed depends on the ``impact parameter" i.e. the distance  from the  star's center.
This will be the case of binaries, and/or occultations of background stars by NS or WD. 
We show in Fig. \ref{fig:prob2} the dependence of the probability conversion 
into axions for two typical values of the coupling energy scale $M$ and typical frequency as a function of  surface magnetic field and radial distance of the transit. This figure is qualitatively different from Fig.\ref{fig:prob1}. 
In fact, when a photon is  generated at the star surface and exits the magnetic field, for a given set of parameters m$_a$ and M,  larger surface magnetic field --above a threshold of $B\sim 10^6-10^8$G --leads to adiabatic photon-axion conversion and therefore to drastic reduction of  dimming. On the other hand for photons crossing the magnetic field of a foreground star, the effect is driven by   the impact parameter. If that is large  enough, the photons do no cross the non-adiabatic region and no conversion is produced, but for moderate surface magnetic fields--$B\sim 10^6$G--, outside the star surface there exist  a region of   non-adiabaticity and therefore sizable conversion.

The two panels of Fig.\ref{fig:prob2} illustrate the sensitivity to the axion-photon coupling in this case:
the sensitivity to m$_a$ is weak for smaller axion masses, while the conversion is lost for much larger values. The results shown in Fig. \ref{fig:prob2}  do not change with frequency in the optical range. 

In the observation of one of such transits, the axion dimming  effect could be seen as a light-curve dimming. Besides statistical errors, the limitation to measure the dimming is  the accuracy of  relative photometry  which can easily be better than \% level. This consideration  can be translated into  the accessible region of the  PQ axion parameter space. \\
{\bf XDINS:}We find that  a factor of few dimming at high frequencies is possible but  for values of the magnetic field $\lesssim10^{8} G$ which are severely in disagreement with the values obtained from the spin-down analysis (about $10^{13} G$). Thus PQ axions 
cannot be the explanation for the observed spectrum of XDINS.  
However, we do find (Fig.1 for $B\sim 10^{12} G$) that if future observations extend to wavelengths beyond the keV range, there could be a noticeable dimming effect even at magnetic fields values as the ones measured for XDINS.  

{\bf    Occultations:}  Let us now investigate the possibility that the effect of photon-axions conversion is observed in the light of a background object passing through the influence of NS  (or WD) magnetic field.  We call this effect ``occultation".  
A simple estimate of  the occultation rate per observed NS ($p'$) is obtained as follows \cite{schwarz}.
Consider a NS at distance $r$ that passes near the line of sight to a background object (which could be a distant galaxy or another galactic star). The occultation rate is proportional to the solid angle swept by the NS which depends on the projected velocity $v$ and the radius from the NS where most of the conversion happens ($R_c$). It also depends on the number density of background objects on the celestial sphere ($n_{bg}$).
Thus we obtain:
\begin{eqnarray}
p'&=&\frac{R_c v}{r^2}n_{\rm bg}\\
&=& 3.3\times 10^{-11} \frac{R_c}{10^3 {\rm km}} \frac{v}{200 {\rm km/s}}\left (\frac{1{\rm kpc}}{r}\right )^2\frac{n_{bg}}{100^{\square'}} \frac{1}{\rm yr} \nonumber
 \end{eqnarray}
where in the second line we have rewritten the equation in terms of typical numbers for  the relevant variables.

There  are $N_{\rm NS}\sim 10^9$ NS in our galaxy, the total occultation rate is therefore obtained integrating over the NS distribution.

We  assume \cite{Narayan1987} that the NS distribution can be factorized as function of the distance from the galactic center $R$  and height over the galactic plane $z$, $n(R,z) 2\pi R dR dz=N_{\rm NS}[n_r(R)2\pi R dR][n_z(z)dz]\,.$
For the $R$ dependence we  consider two cases. The Hartman exponential model:
$n_R(R)=\frac{1}{2\pi R_H^2}\exp[-R/R_H]$, where $R_H=5$kpc; and the Hartmann model  where 
$n_R(R)=\frac{c}{2\pi R^2_w}\exp\left(-\frac{(R-R_{\rm max})^2}{2R_w^2}\right) $
with $R_w=1.8$kpc and $R_{\rm max}=3.5$kpc; $c=0.204$.
For the $z$ dependence  we assume a Gaussian 
%\begin{equation}
%n_z(z)=\frac{1}{\sqrt{2\pi}\sigma}\exp\left(\frac{-z^2}{2\sigma^2}\right)
%\end{equation}
with $\sigma=0.45$kpc \cite{Lyneetal1998}. 
These distributions must be transformed into spherical coordinates ($r$, $\theta$, $\phi$) with the Sun at the origin. By integrating over the $\theta$ and $\phi$ angles we obtain the radial distribution of NS. This can then be used to compute the total occultation rate:
%\begin{eqnarray}
%p&=&\int p' n(r) dr= 0.004-0.02 \\ \nonumber
%&\times&\frac{R_c}{1000 {\rm km}} \frac{v}{200 {\rm km/s}}\frac{n_{bg}}{100^{\square'}} \frac{1}{{\rm yr} } 
%\end{eqnarray}
\begin{equation}
p = \int p' n(r) dr= 0.004-0.02 \times \frac{R_c}{1000 {\rm km}} \frac{v}{200 {\rm km/s}}\frac{n_{bg}}{100^{\square'}} \frac{1}{\rm yr} 
\end{equation}
depending on the radial NS distribution chosen.

\begin{figure}
  \begin{center}
    \includegraphics[height=2in,width=3in]{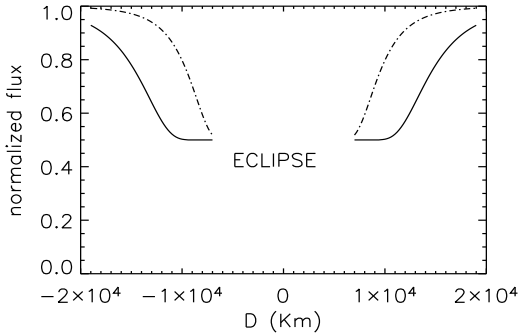}
    \caption{\label{dat_n_BBs} Lightcurve of  an infinitesimal element of the surface of the companion when undergoing eclipse behind a WD. We have assumed a WD radius of $7000$ km, a surface magnetic field of  $10^6$G and the two lines are for $M=10^7$ (solid)  and $3\times 10^7$GeV (dot-dashed).  On the $x$ axis we report the projected distance from the center of the WD.}
  \end{center}
\end{figure}    

Therefore the number of expected occultations from NS is not significant, even if the whole sky is observed. However, we have shown above that  maximum conversion takes place  for surface  magnetic fields of order $10^6$ G. About 10\% of WD have surface magnetic fields of that order or above. In addition, their radius is of the order of  $10^4$ km, which further increases the cross section. There are about $5 \times 10^{10}$ WD in the galaxy. So using eq. (15) we find that for WD occultations (assuming the same spatial distribution in the galaxy as NS) the probability is increased to  $p = 0.1 - 1 yr^{-1}$, which makes the effect much more feasible to be observed.

Surveys that cover the full sky could find that some of their background stars are dimmed when they pass behind a WD at a rate of one per  few years. No survey currently exists that has these characteristics, although in the future Pan-Starrs and LSST \cite{lsst} should be able to provide few events over an operating  time period of a decade. It will be interesting, nevertheless, to explore if some of such events exist in long-running surveys like OGLE. Note that the signature is  achromatic at optical wavelengths, thus distinguishing it from other effects like obscuration.

%\subsection{\lv{White dwarfs in ``eclipsing'' binaries}}
{ \bf WD in ``eclipsing'' binaries}:
Another probe to search for photon-axion conversion is in binaries where one of the pair is a WD or a NS.  As it is estimated that about  60\% of WD are in binaries and WD are much more numerous than NS, we will illustrate here the case for WD.

 Any WD in a binary system with orbital plane close to the line of sight of the observer would  be  an excellent candidate for this phenomenon.  If  the companion's light passes  within the  radius of maximum conversion, $R_C$, dimming  due to photon-axion conversion can be observed.
Consider an infinitesimal element on the surface of the companion star and assume it undergoes a full eclipse, although dimming can happen as long as the background light passes   within $R_C$. In Figure~\ref{dat_n_BBs} we show the light curve for such an element. We assume a typical WD radius of 7000 km, a surface magnetic field of  $10^6$G and the two lines are for $M=10^7$ (solid) and $3\times 10^7$GeV (dashed).  On the $x$ axis we report the projected distance from the center of the WD: this can be mapped into time depending on the orbital speed and geometry.
    In the  WD-WD binary recently discovered 
    \cite{WD-WD}, the  two WD transit in front of each other,  and the dimming effect should appear in the light curves of both primary and secondary eclipse.  
\begin{figure}[t]
  \begin{center}
    \includegraphics[height=2in,width=3in]{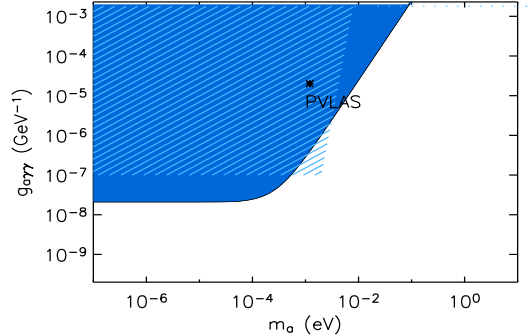}
    \caption{Region in the PQ $m_a$, $g_{a\gamma\gamma}$ plane excluded by the light curve observation of the eclipsing double WD \cite{WD-WD} (dark blue). The hashed light blue region corresponds to  forecasted constraints from binary pulsars observations and the black dot denotes the PVLAS region. We exclude the possibility of explaining the PVLAS result with photon-axion coupling. }
    \label{fig:pq}
  \end{center}
\end{figure}    
 
We fit the phased light curve of the primary eclipse  including the  photon-axion dimming due to the magnetic field of the eclipsing WD.  We keep the  other light curve parameters fixed to the values used in \cite{WD-WD} and assume a magnetic field of 10$^6$ G for the smaller WD. This assumption is motivated by the fact that WD in binaries are accreting WD (formed in a  common envelop) and  thus have higher B \cite{Cumming:2002vq}.
 In practice, the effect illustrated in Fig. 3 is integrated over the  projected surface of the secondary star. We define a $\chi^2$ assuming uncorrelated experimental errors. 
 
 In Fig.~\ref{fig:pq} we show the  resulting 95\% exclusion region in the m$_a$-$g_{a\gamma\gamma}$ parameter space. Using the observed light curves, for the first time, we derive bounds on the photon-axion conversion by the dimming of light in eclipsing binaries. We find that the observations \cite{WD-WD} of the WD eclipsing binary, subject to the assumption we made on the magnetic field strength and the star radius although the region excluded is not very sensitive to these parameters, are able to explore and exclude an interesting  region of axion parameters: this region fully covers the region of direct photon-axion conversion explored  by laser experiments \cite{laser}, excludes the possibility of explaining the PVLAS results with   photon axion conversion \cite{pvlas} and is competitive with the bounds forecasted in proposals like the observation of  high-energy photons in the binary pulsar \cite{doublepulsar} or in high-energy facilities \cite{xfel}. The bounds derived here by analyzing the data of the first WD eclipsing binary depend on the radius and the magnetic field of the WD. These bounds will be robustly determined by the analysis of the data of the large number of eclipsing WD binaries being observed by the ongoing Kepler mission\cite{Kepler}.

%\acknowledgments{
\scriptsize{
We thank T. Avgoustidis for  initial discussions and the hospitality 
and support of IFT-UAM where most of the ideas discussed here were explored.
CPG is supported by the Spanish MICINN grant
FPA-2007-60323 and the Generalitat Valenciana grant PROMETEO/2009/116.
LV acknowledges support from FP7-PEOPLE-2007-4-3-IRG n. 202182 and FP7- 
IDEAS Phys.LSS 240117. LV and  RJ are supported by MICINN grant AYA2008-03531.
%} 
}

\end{document}